\documentclass[12pt]{article}
\pdfoutput=1

\usepackage[bulletsep]{collref}
\usepackage{amssymb,graphicx}
\usepackage[intlimits]{amsmath}
\usepackage{tensor}
\usepackage{dsfont}
\usepackage{bbm}
\usepackage[small]{subfigure}

\usepackage{MnSymbol}

\usepackage{arydshln}

\usepackage{xcolor}

\makeatletter \@addtoreset{equation}{section} \makeatother

\makeatletter
\let\old@startsection=\@startsection
\let\oldl@section=\l@section
\renewcommand{\@startsection}[6]{\old@startsection{#1}{#2}{#3}{#4}{#5}{#6\mathversion{bold}}}
\renewcommand{\l@section}[2]{\oldl@section{\mathversion{bold}#1}{#2}}
\makeatother

\makeatletter
\let\old@makecaption=\@makecaption
\def\@makecaption{\small\old@makecaption}
\makeatother

\newcommand{\kkb}{{\color{teal} \mathring{a}}}
\newcommand{\knb}{{\color{blue} \o}}
\newcommand{\nkb}{{\color{blue} \o}}
\newcommand{\nnb}{{\color{violet} \ae}}
\newcommand{\TT}{\mathbb{S}}

\newcommand{\wrap}{{\rm int}}
\newcommand{\kth}{\mathop{\mathrm{kth}}}
\newcommand{\ktg}{\mathop{\mathrm{ktg}}}

\renewcommand{\a}{\alpha}
\newcommand{\da}{{\dot{\alpha}}}
\newcommand{\db}{{\dot{\beta}}}
\renewcommand{\b}{\beta}
\newcommand{\g}{\gamma}

\newcommand{\dg}{{\dot{\gamma}}}
\newcommand{\Tr}{\mathop{\mathrm{tr}}}

\newcommand{\nn}{\nonumber}

\begin{document}

%%%%%%%%%%%%%%%%%%%%%%%%%%%%%%%%%%%%%%%%%%%%%%%%%%%%%%%%%%%%%%%%%%%%%%%%%%%%%%%%

\thispagestyle{empty}
\begin{flushright}\footnotesize
%\texttt{ITEP-TH-nn/yy}\\
\texttt{NORDITA-2020-041} 
\vspace{0.6cm}
\end{flushright}

\renewcommand{\thefootnote}{\fnsymbol{footnote}}
\setcounter{footnote}{0}

\begin{center}
{\Large\textbf{\mathversion{bold} Integrable boundary states 
in D3-D5 dCFT:
  \\
 beyond scalars}
\par}

\vspace{0.8cm}

\textrm{Charlotte~Kristjansen$^{1}$, Dennis~M\"uller$^{1}$ and
Konstantin~Zarembo$^{1,2}$\footnote{Also at ITEP, Moscow, Russia}}
\vspace{4mm}

\textit{${}^1$Niels Bohr Institute, Copenhagen University, Blegdamsvej 17, 2100 Copenhagen, Denmark}\\
\textit{${}^2$Nordita, KTH Royal Institute of Technology and Stockholm University,
Roslagstullsbacken 23, SE-106 91 Stockholm, Sweden}\\
\vspace{0.2cm}
\texttt{kristjan@nbi.dk, dennis.muller@nbi.ku.dk, zarembo@nordita.org}
%\vspace{3mm}

\vspace{3mm}

%%%%%%%%

\par\vspace{1cm}

\textbf{Abstract} \vspace{3mm}

\begin{minipage}{13cm}
A D3-D5 intersection gives rise to a defect CFT, wherein the rank of the gauge group jumps by $k$ units across a domain wall. The one-point functions of local operators in this set-up map to overlaps between on-shell Bethe states in the underlying spin chain and a boundary state representing the D5 brane. Focussing on the $k=1$ case, we extend the construction to gluonic and fermionic sectors, which was prohibitively difficult for $k>1$. As a byproduct, we test an all-loop proposal for the one-point functions in the $su(2)$ sector at the half-wrapping order of perturbation theory.  
\end{minipage}
\end{center}

\vspace{0.5cm}

%%%%%%%%%%%%%%%%%%%%%%%%%%%%%%%%%%%%%%%%%%%%%%%%%%%%%%%%%%%%%%%%%%%%%%%%%%%%%%%%

%\newpage
\setcounter{page}{1}
\renewcommand{\thefootnote}{\arabic{footnote}}
\setcounter{footnote}{0}

%%%%%%%%%%%%%%%%%%%%%%%%%%%%%%%%%%%%%%%%%%%%%%%%%%%%%%%%%%%%%%%%%%%%%%%%%%%%%%%%

\section{Introduction}

 The study  of supersymmetric boundary conditions in ${\cal N}=4$ SYM is mathematically  
enriching~\cite{Gaiotto:2008sa,Gaiotto:2008ak} and provides new avenues for the development of exact methods 
for extracting the theory's observables, such as integrability~\cite{Beisert:2010jr}, localization~\cite{Pestun:2016zxk} and the boundary conformal bootstrap program~\cite{Liendo:2012hy,Liendo:2016ymz}.

Domain wall versions of ${\cal N}=4$ SYM with gauge groups of different rank  on the two sides of the wall provide examples of set-ups for which the boundary conditions can be chosen to conserve half of the supersymmetries. More specifically by
assigning an expectation value in the form of a Nahm pole to three of the scalars of ${\cal N}=4$ SYM for, say $x_3>0$, one
can arrive at a situation where the gauge group is $SU(N+k)$ for $x_3>0$ and $SU(N)$ for $x_3<0$~\cite{Nahm:1979yw,Diaconescu:1996rk,Karch:2000gx}. This construction
is formally restricted to $k>1$ whereas the dual string theory set-up which corresponds to having a single D5 brane being the
end locus for  $(N+k)$ D3-branes for $x_3\rightarrow 0_+$ and $N$ D3-branes for $x_3\rightarrow 0_-$~\cite{Karch:2000gx,Constable:1999ac} does not seem to 
entail a similar restriction. In the present paper we present a thorough field theoretical analysis of the $k=1$ case showing that
indeed results for quantum observables  obtained for $k>1$ apply directly to the  $k=1$ case as well.  We mention in passing
that there exists another 1/2 BPS defect version of ${\cal N}=4$ SYM, not of domain wall type, which can be said to correspond
to $k=0$~\cite{DeWolfe:2001pq}. For this set-up there are no boundary conditions on the bulk fields but an additional fundamental hypermultiplet lives on the defect.  The relation between the $k=0$ and the $k=1$ cases was analyzed in~\cite{Ipsen:2019jne},
where it was argued that $k=1$ should be the simplest dCFT from the integrability point of view. We can say, skipping ahead, that our findings fully confirm this assertion. 

 Integrability has lead to significant progress in determining quantum observables in  domain wall versions of ${\cal N}=4$ SYM of the type above with the focus being on one-point functions, the simplest observables of defect CFTs. Initiated by the derivation of an exact formula for tree-level one-point functions of the $SU(2)$ sector 
 in~\cite{deLeeuw:2015hxa,Buhl-Mortensen:2015gfd}, the analysis was extended
 to the full scalar  sector~\cite{deLeeuw:2016umh,deLeeuw:2018mkd} and to one-loop order 
 in~\cite{Buhl-Mortensen:2016pxs,Buhl-Mortensen:2016jqo}.  The one-loop analysis moreover lead to the
 conjecture of an exact expression for an all loop asymptotic one-point function formula for the $SU(2)$ sub-sector~\cite{Buhl-Mortensen:2017ind}.  
 
Recent bootstrap solution for the boundary state \cite{Komatsu:2020sup}, based on the ideas from \cite{Jiang:2019xdz,Jiang:2019zig}, opens an avenue to study one-point functions in dCFT at a fully non-perturbative level. Having confirmed the asymptotic $SU(2)$ formula, this approach can potentially be extended to all operators and is capable to incorporate finite-size effects through a TBA-like formalism  \cite{Jiang:2019xdz,Jiang:2019zig}. One-point functions of protected operators, non-trivial in dCFT,  can be efficiently computed by localization \cite{Robinson:2017sup,Wang:2020seq}, as shown in \cite{Komatsu:2020sup}.

Translation of perturbation theory into the spin-chain language is the stepping stone to advanced methods of integrability, and we will analyze the one-point functions in dCFT in this vein, focusing on the $k=1$ case mostly ignored in previous analyses. 
As we shall see, the intrinsic simplicity of the $k=1$ interface
provides an easy trajectory for going beyond the scalar sector, addressing both gauge fields and fermions.

 Our paper is organized as follows.  We start by describing  the domain wall boundary conditions of ${\cal N}=4$ SYM both for   $k>1$ and $k=1$ in section~\ref{Bc}. Subsequently, we review in section~\ref{Scalars} the integrability properties of one-point functions in the $SU(2)$ sub-sector for $k>1$. In particular, we present the conjectured asymptotic all loop formula and show
that it gives a perfectly meaningful description of the $k=1$ case as well, predicting the one-point function to start out at loop
order $\lambda^{L/2}$. In section~\ref{SU2k=1} we explicitly compute the leading order contribution and find perfect agreement with the asymptotic prediction. The extension of our perturbative computation to the full scalar sector is immediate and is described in section~\ref{SO(6)}.
In the subsequent sections, we show how the $k=1$ domain wall model allows us to easily access one-point functions in other sectors, namely a sector containing purely gluonic operators  as well as sectors containing fermions.
More specifically, we obtain a closed leading order formula for one-point functions in the sector consisting of self-dual 
field strengths in section~\ref{Gluons} and in the simplest sector containing fermions, $SU(2|1)$, in section~\ref{Fermions}. 
Novel tools to compute integrable overlaps
\cite{Pozsgay:2018ixm,Piroli:2018ksf,deLeeuw:2019ebw,Gombor:2020kgu} will be instrumental for our analysis.
We have relegated a discussion of our conventions for spinors as well as a detailed discussion of the factorization properties of  Gaudin determinants to appendices. Finally, section~\ref{Conclusion} contains our conclusion.

\section{D3-D5 dCFT \label{Bc}}

\begin{figure}[t]
\begin{center}
 \centerline{\includegraphics[width=8cm]{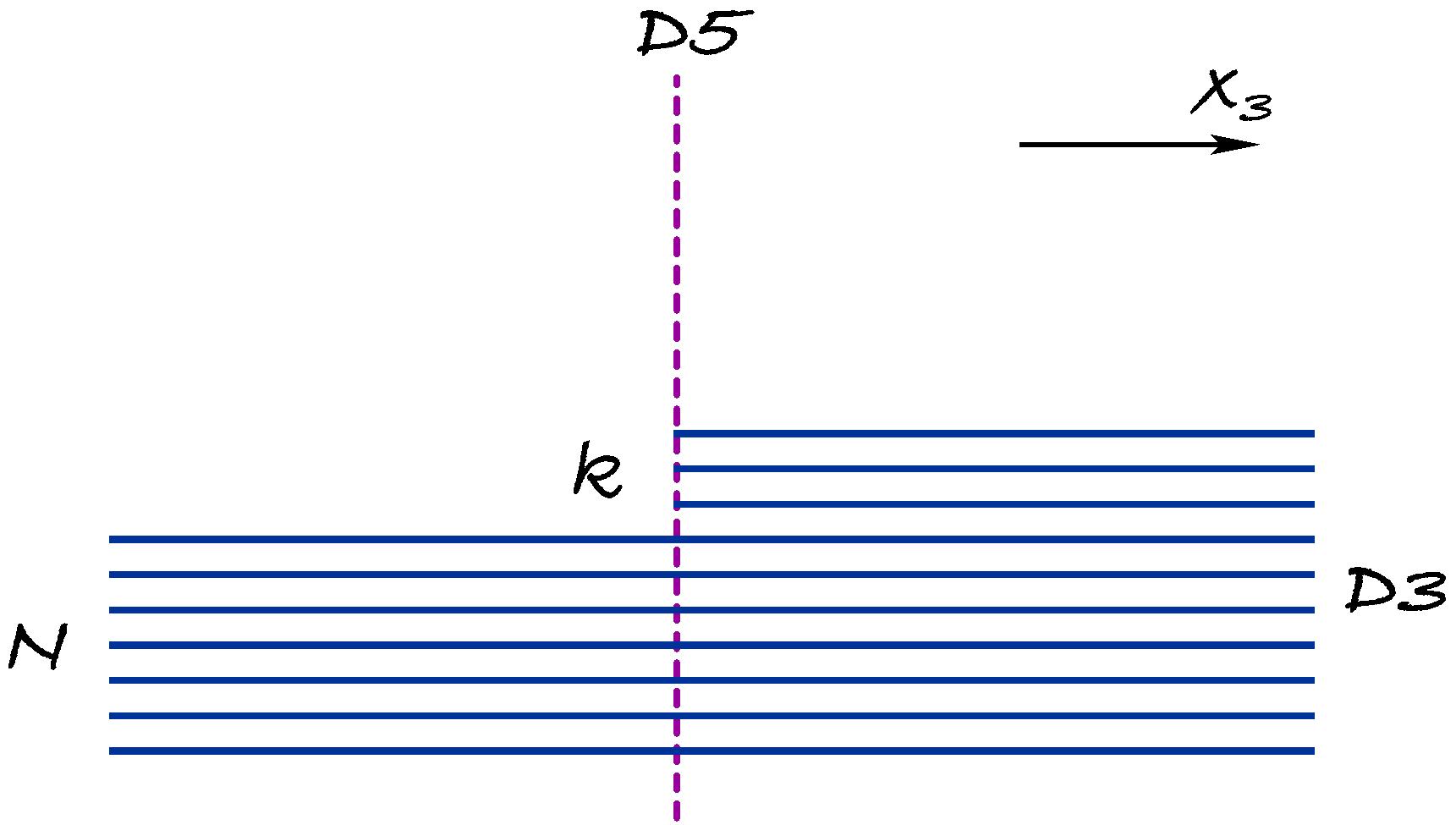}}
\caption{\small The D3-D5 dCFT.\label{D3D5-fig}}
\end{center}
\end{figure}

The matter content of the $\mathcal{N}=4$ theory consists of six adjoint scalars $\Phi _i$ and four adjoint fermions $\Psi $:
\begin{equation}
 \mathcal{L}=\frac{1}{g^2_{\rm YM}}
 \mathop{\mathrm{tr}}\left\{
 -\frac{1}{2}\,F_{\mu \nu }^2+\left(D_\mu \Phi _i\right)^2+\frac{1}{2}\,\left[\Phi _i,\Phi _j\right]^2
+i\bar{\Psi }\gamma ^\mu D_\mu \Psi 
 +\bar{\Psi }\gamma ^i\left[\Phi _i,\Psi \right]
 \right\}
\end{equation}
The fermions can be packaged into a single ten-dimensional Majorana-Weyl spinor, so that the gamma matrices $(\gamma  ^\mu ,\gamma  ^i)$ form a representation of the ten-dimensional Clifford algebra which can be realized by supplementing $\gamma  ^\mu $ with $\gamma ^i=\gamma ^5\Gamma ^i$, where the 6D Dirac matrices $\Gamma ^i$ act exclusively on the R-symmetry indices. The Majorana-Weyl conditions require $\gamma ^{11}\Psi =\Psi $ and $\bar{\Psi }=\Psi ^tC$, where $C$ is the 10d charge conjugation.

The domain wall defect we are going to discuss separates regions of spacetime with different gauge groups: $U(N)$ at $x_3<0$ and $U(N+k)$ at $x_3>0$. For the time being we  keep $k$ arbitrary. The set-up is illustrated in fig.~\ref{D3D5-fig} and is accomplished by a block decomposition of the fields according to the symmetry-breaking pattern:
\begin{equation}\label{N+k-decomp}
\begin{array} {ll}
\hphantom{A_\mu ,\Phi _i,\Psi  = }~~~~\!k\,\,\,~~~~~~N &  \\[1pt]
   A_\mu ,\Phi _i,\Psi= \left[
 \begin{array}{c:ccc}
 \kkb & \knb & \knb & \knb \\
 \hdashline
  \nkb & \nnb  & \nnb & \nnb \\
   \nkb & \nnb & \nnb & \nnb  \\
    \nkb & \nnb & \nnb & \nnb \\
 \end{array}
 \right]
 \begin{array}{c}
 k\\
   \\
  N  \\
    \\
 \end{array}
  \end{array}
\end{equation}
The $\nnb$-fields propagate in the whole space, while the $\kkb$ and $\knb$ components  are confined to $x_3>0$.

The confinement mechanism is markedly different at $k>1$ and at $k=1$, or so it looks at the first sight. When $k>1$, the scalars acquire vacuum expectation values \cite{Diaconescu:1996rk,Gaiotto:2008sa}:
\begin{equation}\label{phiclass}
 \Phi _i^{\rm cl}=\frac{t_i}{x_3}\,,~i=1,2,3;\qquad \Phi _i^{\rm cl}=0,~i=4,5,6;
\end{equation}
where the $k\times k$ matrices $t_i$ are restricted to the $\kkb$ block and satisfy the $\mathfrak{su}(2)$ commutation relations $[t_i,t_j]=i\varepsilon _{ijk}t_k$.  The classical background breaks color as prescribed and in addition reduces the $SO(6)$ R-symmetry to $SO(3)\times SO(3)$. The Higgs mechanism induces space-varying masses $m^2\propto 1/x_3^2$ for  the  fields in the $\kkb$ and $\knb$ blocks, which grow without bound as $x_3\rightarrow 0^+$. The resulting potential barrier repels those fields from the $x_3<0$ domain effectively confining them to one side of the domain wall.

There is no classical background for $k=1$, so the $\kkb$ and $\knb$ fields have to be restricted to the half-space by hand, by imposing generalized Neumann/Dirichlet boundary conditions at $x_3=0$: 
\begin{eqnarray}\label{bcs}
 &D_3\Phi _i+\frac{i}{2}\,\varepsilon _{ijk}[\Phi _j,\Phi _k]=0, &i=1,2,3,
\nonumber \\ 
&\Phi _i=0, &i=4,5,6,
\nonumber \\
&F_{\mu \nu }=0,&\mu ,\nu =0,1,2,
\nonumber \\
&i\gamma ^3\Psi =\Psi, &
\end{eqnarray}
very much in line with the structure of the D-brane intersection.
Indeed, a semi-infinite D3 brane can slide along the D5 brane, hence Neumann conditions in the $i=1,2,3$ directions, but cannot split from it, hence Dirichlet for $i=4,5,6$. The rest follows from supersymmetry  \cite{Gaiotto:2008sa}.

At the quantum level the apparent differences between $k=1$ and $k>1$ should disappear. After all, the symmetry breaking pattern is the same in both cases. One manifestation of similarity between $k>1$ and $k=1$ is this: setting formally $k=1$ in the propagators around the classical background (\ref{phiclass}) \cite{Buhl-Mortensen:2016jqo} one gets the standard Dirichlet/Neumann Green's functions. We expect that correlation functions are in some sense "analytic" in $k$ and to draw some intuition about $k=1$ we start by reviewing the much better understood  $k>1$ case.

\section{One-point functions: scalars \label{Scalars}}

One-point functions in dCFT are fixed by scale invariance up to a constant:
\begin{equation}
 \left\langle \mathcal{O} (x)\right\rangle=\frac{C_{\mathcal{O}}}{x_3^\Delta }\,.
\end{equation}
The constant carries dynamical information both about the operator and the defect, and is unambiguous once $\mathcal{O}$ is properly normalized, for instance by its two-point function at asymptotic infinity: we assume that $\left\langle \bar{\mathcal{O}} (x)\mathcal{O}(y)\right\rangle\simeq 1/|x-y|^{2\Delta }$ at $x_3,y_3\rightarrow \infty $. Conformal boosts impose additional constraints, in particular, one-point functions of conformal primaries with a non-zero Lorentz spin must vanish \cite{Liendo:2012hy}.

The tree-level one-point functions for $k>1$ are obtained by simply setting all the fields in the operator to their classical values. Take, for instance, a generic scalar operator
\begin{equation}
 \mathcal{O}=\Psi ^{i_1\ldots i_L}\mathop{\mathrm{tr}}\Phi _{i_1}\ldots \Phi _{i_L}.
\end{equation}
A cyclically symmetric tensor $\Psi ^{i_1\ldots i_L}$ can be interpreted as a wavefunction in an integrable  spin chain of length $L$, with the vector representation of $SO(6)$ at each site. The one-point function is then represented by an overlap of the operator's wavefunction with a fixed external state:
\begin{equation}\label{structure-const}
 C_{\Psi }=\left(\frac{4\pi ^2}{\lambda }\right)^{\frac{L}{2}}L^{-\frac{1}{2}}\,
 \frac{\left\langle B\right.\!\left|\Psi  \right\rangle}{\left\langle \Psi \right.\!\left|\Psi  \right\rangle^{\frac{1}{2}}}\,.
\end{equation}
The overall prefactor accounts for a difference in the spin-chain and field-theory normalizations: we use the complex field conventions:
\begin{equation}
 Z=\Phi _1+i\Phi _4\,,\qquad X=\Phi _2+i\Phi _5\,,\qquad Y=\Phi _3+i\Phi _6,
\end{equation}
and normalize $\left\langle Z\right.\!\left| Z\right\rangle=1$ and so on.

\begin{figure}[t]
\begin{center} 
 \subfigure[]{
   \includegraphics[height=1.5cm] {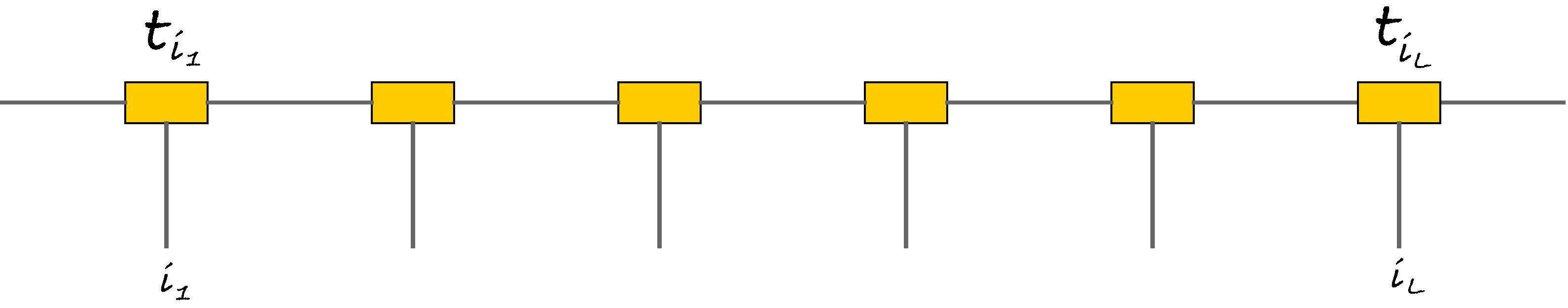}
   \label{MPS-subfig}
 }
 \subfigure[]{
   \includegraphics[height=1.5cm] {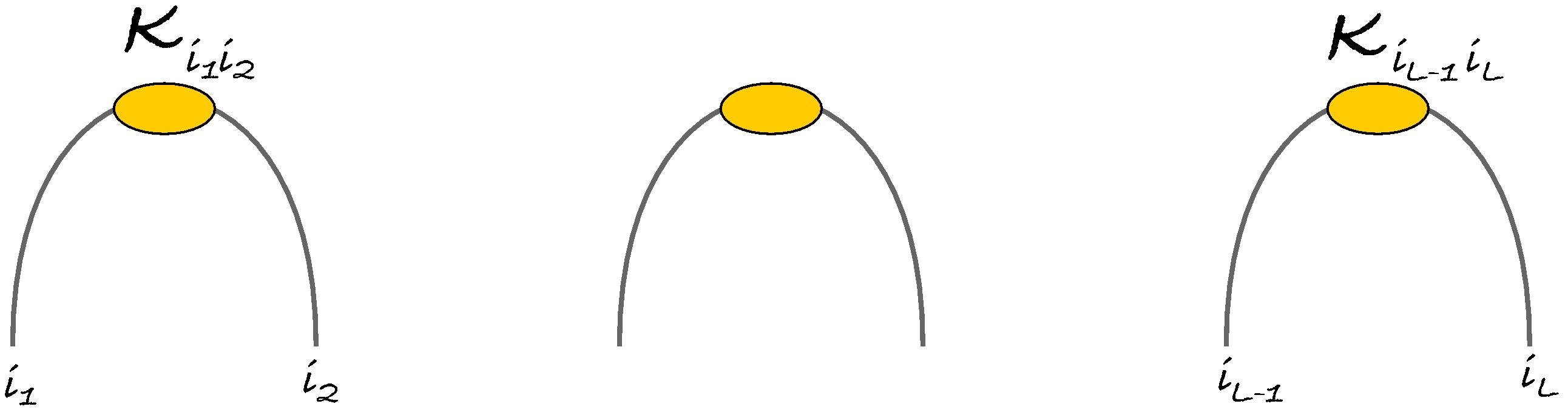}
   \label{VB-subfig}
 }
\caption{\label{MPSvsVB}\small 
Two types of integrable boundary states:
(a) MPS (b) VBS. }
\end{center}
\end{figure}

For the operators at hand the external wavefunction takes the form of a Matrix Product State (MPS) \cite{deLeeuw:2015hxa}:
\begin{equation}\label{MPS-gen}
 B_{i_1\ldots i_L}=\mathop{\mathrm{tr}}t_{i_1}\ldots t_{i_L}.
\end{equation}
Crucially, this external state is integrable, meaning that spin-chain magnons appear in its wavefunction in momentum-conjugate pairs $\left\{p,-p\right\}$. This is the standard criterion for boundary integrability 
\cite{Ghoshal:1993tm}, 
elaborated in detail in \cite{Piroli:2017sei}, a direct counterpart of reflection elasticity in  the cross channel. 
Integrable boundary states have never been fully classified but, by experience, come in two broad categories. One is MPS as above \cite{Pozsgay:2018dzs}, but more conventional boundary states are associated with reflection matrices and have a Valence Bond (VB) structure:
\begin{equation}\label{VBS-gen}
 B^{\rm VBS}_{i_1\ldots i_L}=K_{i_1i_2}\ldots K_{i_{L-1}i_L}.
\end{equation}
The difference between MPS and VBS is illustrated in fig.~\ref{MPSvsVB}.

Integrable boundary states have remarkably simple overlaps with the on-shell Bethe states. The overlap formulas always have the same architecture, which we illustrate on a simple example, the $\mathfrak{su}(2)$ sector composed of operators $\mathop{\mathrm{tr}}Z^{L-M}X^M+{\rm permutations}$. Their mixing is described by the Heisenberg model:
\begin{equation}\label{Heisenberg}
 H=\sum_{l=1}^{L}\left(1-P_{l,l+1}\right),
\end{equation}
where $P_{l,l+1}$ permutes spins on $l$th and $(l+1)$th sites (fields at $l$th and $(l+1)$th positions inside the trace). The eigenstates are conformal operators with conformal dimension $L+\lambda E/16\pi ^2+\mathcal{O}(\lambda ^2)$, where $\lambda =g_{\rm YM}^2N$ is the 't~Hooft coupling.

The Bethe ansatz solution of the Heisenberg model assigns a set of $M$ Bethe roots $\mathbf{u}=\left\{u_1,\ldots ,u_M\right\}$ to each eigenstate. The roots must satisfy Bethe Ansatz Equations (BAE):
\begin{equation}\label{BAE-su2}
 \,{\rm e}\,^{i\chi_j}\equiv \left(\frac{u_j-\frac{i}{2}}{u_j+\frac{i}{2}}\right)^L
 \prod_{k}^{}\frac{u_j-u_k+i}{u_j-u_k-i}=-1.
\end{equation}
The energy and momentum of the state are given by
\begin{equation}\label{EP-su2z}
 E_{\mathbf{u}}=\sum_{j}^{}\frac{2}{u_j^2+\frac{1}{4}}\,,\qquad 
 \,{\rm e}\,^{iP_{\mathbf{u}}}=\prod_{j}^{}\frac{u_j+\frac{i}{2}}{u_j-\frac{i}{2}}\,.
\end{equation}
Due to trace cy\-cli\-ci\-ty only zero-momentum states correspond to SYM operators. For paired states the momentum constraint is automatic.

The $\mathfrak{su}(2)$ counterpart of (\ref{MPS-gen}) is built out of two $k\times k$ matrices $t_1$ and $t_2$, which are just Pauli matrices  for $k=2$. Since $\sigma _1$, $\sigma _2$ can be rotated to $\sigma _+$, $\sigma _-$, the $k=2$ MPS is unitary equivalent to the N\'eel state, a VBS (\ref{VBS-gen}) with $K=\uparrow\downarrow$ \cite{Piroli:2017sei}. This, along with a relation between MPS$_{k+2}$  and MPS$_k$ \cite{Buhl-Mortensen:2015gfd}, reduces one-point functions to N\'eel overlaps of the on-shell eigenstates of the Heisenberg Hamiltonian, an explicit determinant representation for which  \cite{Brockmann:2014a,Brockmann:2014b,Brockmann2014} can be obtained by a limiting procedure from an off-shell formula \cite{Pozsgay:2009} related by crossing to the partition function of the six-vertex model with reflective boundary
\cite{Tsuchiya:qf}.

The key ingredient of the N\'eel overlaps \cite{Brockmann:2014a,Brockmann:2014b,Brockmann2014}, and of all other known overlap formulas, is the factorized Gaudin matrix. The Gaudin matrix itself is the Jacobian 
\begin{equation}
 G_{jk}=\frac{\partial \chi _j}{\partial u_k}\,,
\end{equation}
where $\chi _j$ is defined in (\ref{BAE-su2}).
Integrability selection rules pick only states with paired rapidities:
$\mathbf{u}=\left\{u_j,-u_j\right\}_{j=1\ldots M/2}$. The Gaudin matrix in this case has a $2\times 2$ block structure and its determinant factorizes:
\begin{equation}
 \det G=\det G^+\det G^-,
\end{equation}
essentially due to antisymmetry of the S-matrix under momentum flip.
Explicitly, for the Heisenberg model,
$$
 G^\pm_{jk}=K^\pm_{jk}+\delta _{jk}\left(\frac{L}{u_j^2+\frac{1}{4}}-\sum_{l}^{}K^+_{jl}\right),\qquad 
 K^\pm_{jk}=\frac{2}{(u_j-u_k)^2+1}\pm\frac{2}{(u_j+u_k)^2+1}\,.
$$   

The overlap can be expressed through the ratio of the Gaudin factors and the Baxter polynomial:
\begin{equation}
 Q(u)=\prod_{j=1}^{\frac{M}{2}}\left(u^2-u_j^2\right),
\end{equation}
evaluated as specific values of the argument.
For the rank-$k$ MPS \cite{Buhl-Mortensen:2015gfd}:
\begin{equation}\label{MPSk-u}
 \frac{\left\langle B_k\right.\!\left|\mathbf{u} \right\rangle}{\left\langle \mathbf{u}\right.\!\left|\mathbf{u} \right\rangle^{\frac{1}{2}}}
 =\TT_{k}Q\left(\frac{ik}{2}\right)\sqrt{Q\left(\frac{i}{2}\right)Q\left(0\right)\,\frac{\det G^+}{\det G^-}}\,,
\end{equation}
where $\TT_k$ is related to the transfer-matrix eigenvalue in the $k$-dimensional representation:
\begin{equation}\label{Tk}
 \TT_k=\sum_{a=-\frac{k-1}{2}}^{\frac{k-1}{2}}\frac{a^L}{Q\left(\frac{2a+1}{2}\,i\right)Q\left(\frac{2a-1}{2}\,i\right)}\,.
\end{equation}

We mention in passing that the ratio of determinants in (\ref{MPSk-u}) is the superdeterminant of the Gaudin matrix with respect to the $\mathbbm{Z}_2$ parity that acts by interchanging paired roots: $\Omega : u_j\rightarrow -u_j$:
\begin{equation}
 \frac{\det G^+}{\det G^-}=\mathop{\mathrm{Sdet}}\nolimits_\Omega  G.
\end{equation}
It is unclear if this representation has any practical advantage\footnote{perhaps if combined with the free-field construction of \cite{Kostov:2019sgu}.}, but it gives a more invariant definition of the overlap that {\it a  priori} does not rely on any decomposition of the Gaudin matrix. 

How do quantum effects change the one-point functions? Higher-loop corrections deform the spin-chain Hamiltonian and consequently its eigenstates $\left|\mathbf{u}\right\rangle$. The boundary state $\left\langle B\right|$ also receives quantum corrections, systematically calculable in perturbation theory whose complexity grows very fast due the non-trivial background field. Current state of the art is one loop   \cite{Buhl-Mortensen:2016pxs,Buhl-Mortensen:2016jqo,Grau:2018keb,Gimenez-Grau:2019fld}.

But as far as the final overlap formula is concerned, all its ingredients are known non-per\-tur\-ba\-tive\-ly, to all orders in the 't~Hooft coupling. The Gaudin matrix of the asymptotic BAE \cite{Beisert:2005fw}, an all-loop counterpart of (\ref{BAE-su2}), factorizes for the paired states, the Baxter polynomial and the transfer matrix can be naturally generalized to higher loops.
Taking this into account, an all-loop overlap formula  was conjectured in \cite{Buhl-Mortensen:2017ind}, and was recently derived by bootstrap methods along with the requisite dressing factors \cite{Komatsu:2020sup}. Akin to the all-loop BAE, the overlap formula is asymptotic, valid up to the wrapping order.

Quantum corrections affect the transfer matrix in the following way:
\begin{equation}\label{all-loop-Tk}
  \TT_k^{\rm all-loop}=\sum_{a}^{}\frac{x^L_a\sigma _a}{Q\left(\frac{2a+1}{2}\,i\right)Q\left(\frac{2a-1}{2}\,i\right)}\,.
\end{equation}
The quantum-deformed spin labels $x_a$ are defined by the Zhukovsky formula:
\begin{equation}\label{Zhuk}
 2x_a=a+\sqrt{a^2+\frac{\lambda }{4\pi ^2}}\,.
\end{equation}
The dressing factors $\sigma _a$ depend on the Bethe roots, much like the Baxter polynomial. Their exact functional form can be found in \cite{Komatsu:2020sup}, but for our purposes it will suffice to know that at tree level they trivialize: $\sigma _a=1+\mathcal{O}(\lambda )$.

The summation range in (\ref{all-loop-Tk}) is $a=-\frac{k-1}{2}\ldots \frac{k-1}{2}$ for $k$ even and $a=-\frac{k-1}{2}\ldots -0,+0\ldots \frac{k-1}{2}$ for $k$ odd \cite{Buhl-Mortensen:2017ind}. The $a=0$ term is counted twice with two regularizations that shift $x_a$ to the left and to the right of the Zhukovsky cut.

Having sketched the one-point functions at $k>1$, we may now ask what happens if  $k$ is set to one. At tree level we find, correctly, that the overlap is zero, because only the $a=0$ term remains in the sum, and in (\ref{Tk}) this term is zero. But in the all-loop character (\ref{all-loop-Tk}) the two regularized $a=0$ terms survive, albeit are hugely suppressed at weak coupling. To the first non-vanishing order,
\begin{equation}\label{B1}
 \frac{\left\langle B_1\right.\!\left|\mathbf{u} \right\rangle}{\left\langle \mathbf{u}\right.\!\left|\mathbf{u} \right\rangle^{\frac{1}{2}}}
 =2\left(\frac{\lambda }{16\pi ^2}\right)^{\frac{L}{2}}
 \sqrt{\frac{Q\left(0\right)}{Q\left(\frac{i}{2}\right)}\,\,\frac{\det G^+}{\det G^-}}\,.
\end{equation}

This line of reasoning suggests that the one-point functions for $k=1$ start at $L/2$ loops. The predicted determinant representation is a typical formula for  integrable overlaps in the $\mathfrak{su}(2)$ spin chain.
Such a simple result should have an equally simple explanation.

\subsection{$SU(2)$ sector at $k=1$\label{SU2k=1}}

To check this prediction we are going to compute one-point functions directly, by quantizing the theory with the boundary conditions (\ref{bcs}). These are either Neumann or Dirichlet or no boundary conditions at all, depending on the field component. They are summarized in the following chart:
\begin{equation}
\begin{tabular}{c|c|c} & $\Phi_{4,5,6}, A_{0,1,2}, c$ &  $\Phi_{1,2,3}, A_3$ \\ \hline
$\knb, \kkb$ &  Dirichlet  & Neumann \\
$\nnb $ & no BCs & no BCs  
\end{tabular}
\end{equation}
The scalar propagator for all three types of boundary conditions is given by a single formula
\begin{equation}
 D_\kappa (x,y)=\frac{1}{4\pi ^2}\left(\frac{1}{|x-y|^2}+\frac{\kappa }{|\bar{x}-y|^2}\right),\
\end{equation}
where $\bar{x}=(x_0,x_1,x_2,-x_3)$ and
\begin{equation}
 \kappa =
\begin{cases}
 1 & {\rm Neumann}
\\
  -1 & {\rm Dirichlet}
  \\
  0 & {\rm no~BCs}.
\end{cases}
\end{equation}

Our goal is to calculate $\left\langle \mathcal{O}(x)\right\rangle$ for
\begin{equation}
 \mathcal{O}=\Psi ^{s_1\ldots s_L}\mathop{\mathrm{tr}}Z_{s_1}\ldots Z_{s_L},
\end{equation}
where $s_l=\uparrow,\downarrow$; $Z_\uparrow=Z$, and $Z_\downarrow=X$. To the lowest order in perturbation theory the fields in the operator are  Wick contracted among themselves. Each $Z_s$ is an $(N+1)\times (N+1)$ matrix decomposed as in (\ref{N+k-decomp}). The fields in the $N\times N$ block do not contribute, because the conventional propagator vanishes for chiral fields: $\left\langle Z_s^{ab}Z^{cd}_r\right\rangle=0$ for $a,b,c,d=2\ldots N+1$. On the contrary,  the $\knb$ and $\kkb$ components have non-trivial propagators even for the fields of the same chirality:
\begin{equation}
 \left\langle Z_s^{1a}(x)Z_r^{b1}(y)\right\rangle=\frac{g_{\rm YM}^2\delta _{rs}\delta ^{ab}}{2}\Bigl(D_1(x,y)-D_{-1}(x,y)\Bigr)=
 \frac{g_{\rm YM}^2\delta _{rs}\delta ^{ab}}{4\pi ^2|\bar{x}-y|^2}\,,
\end{equation}
just because real and imaginary parts of $Z^{1a}$ satisfy different boundary conditions. 

\begin{figure}[t]
\begin{center} 
 \subfigure[]{
   \includegraphics[height=4.5cm] {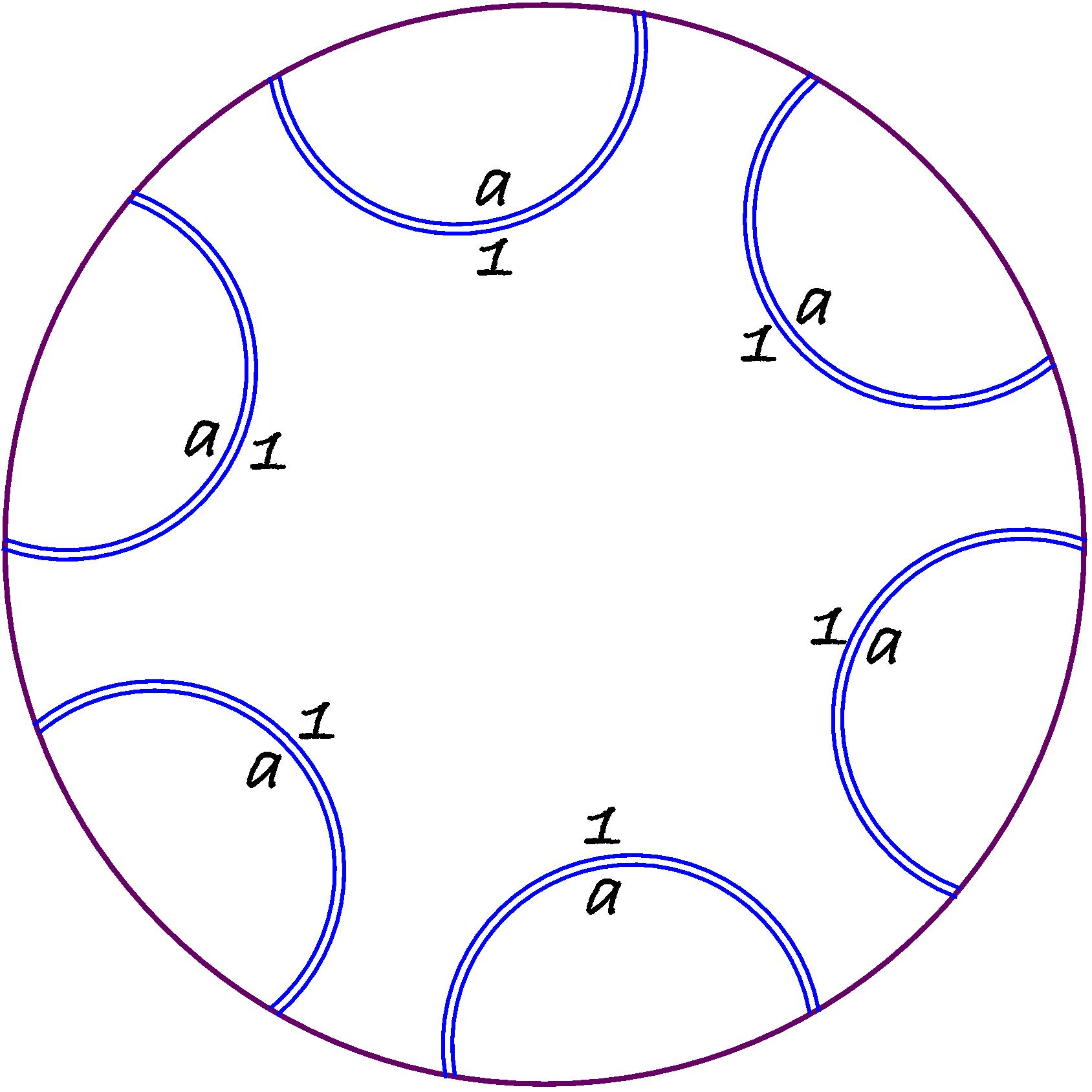}
 }
 \subfigure[]{
   \includegraphics[height=4.5cm] {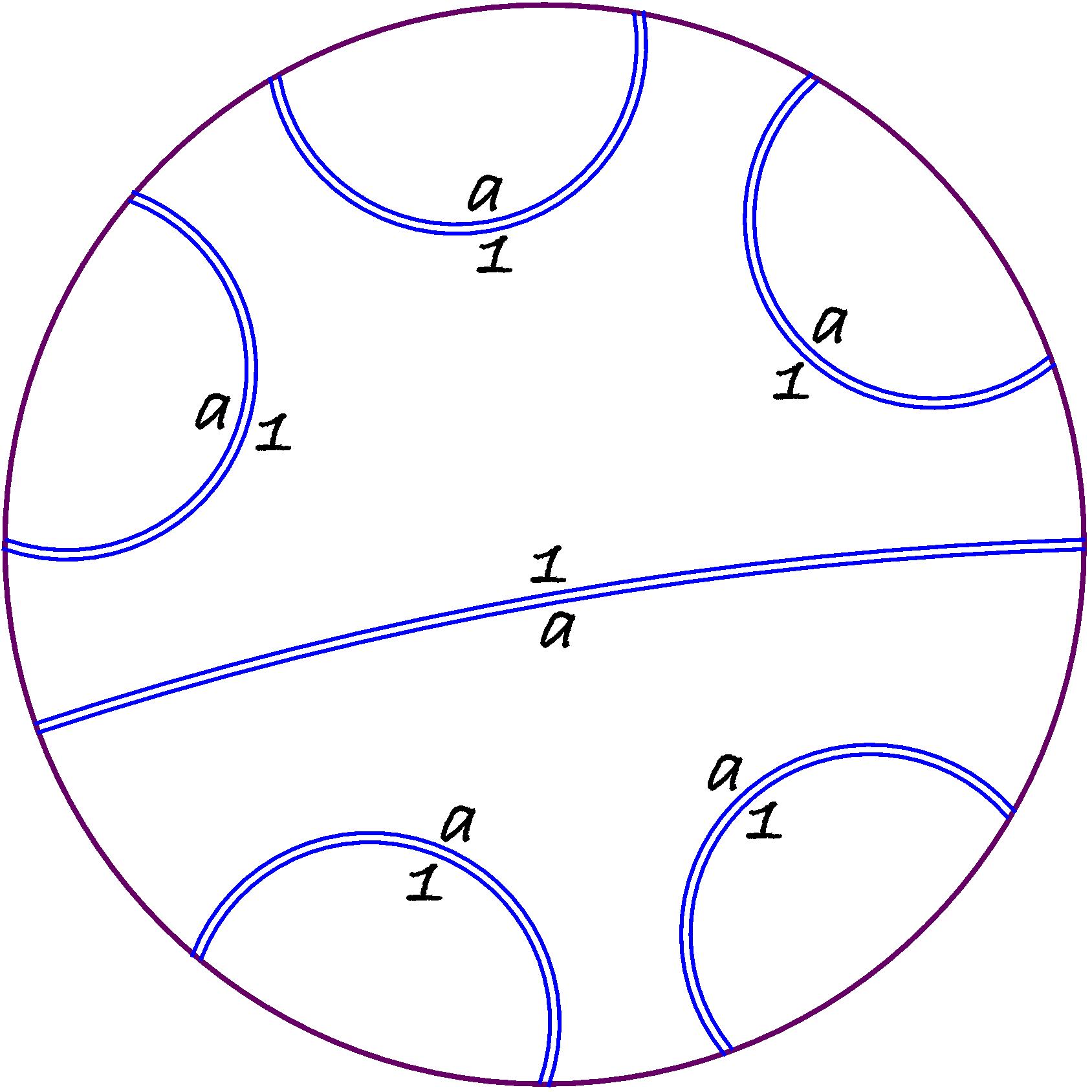}
 }
\caption{\label{Diagrams}\small 
Diagrams that contribute to the one-point function. The operator is depicted as a circle, to visualize the color trace. In spacetime the operator is a point, and all the propagators produce the vacuum bubble $D(0)$. Difference in the index structure allocates the two diagrams to different orders in $1/N$:
(a) The leading-color contribution, in this case  $\mathcal{O}(N^6)$. (b) A subleading diagram, contributing at $\mathcal{O}(N^4)$.}
\end{center}
\end{figure}

The chiral propagator is non-singular at coincident points and the product of $L/2$ propagators produces the requisite $1/x_3^L$ factor. The computation thus reduces to simple combinatorics, which is further simplified by planarity. In the double-line notation, one index of each line must be $1$, the other can be $1$ or can be an unconstrained index to be summed over. We want to maximize the number of unconstrained index loops. It is easy to see that the maximum is achieved by contracting nearest neighbors, as illustrated in fig.~\ref{Diagrams}. The color factor of the leading diagram is $N^{L/2}$, neatly combining with $g^L_{\rm YM}$ to an overall factor of $\lambda ^{L/2}$.

The flavor indices of the nearest neighbors get identified, and we conclude that the one-point function, to the leading order in perturbation theory, is given by an overlap with the VBS-type boundary state:
\begin{equation}
 C_{\mathbf{u}}=2^{-L}L^{-\frac{1}{2}}\frac{\left\langle {\rm VBS}_K\right|
(1+U)
 \left|\mathbf{u}\right\rangle}{\left\langle \mathbf{u}\right.\!\left|\mathbf{u} \right\rangle^{\frac{1}{2}}}\,,
\end{equation}
the latter defined by an elementary two-site block
\begin{equation}\label{uudd}
 K_{sr}=\delta _{sr},\qquad \left\langle K\right|=\left\langle\uparrow\uparrow\right|+\left\langle\downarrow\downarrow\right|.
\end{equation}
The translation operator $U$ accounts for the two possible contractions. Since the operators are cyclically symmetric, $(1+U)$ can be simply replaced by $2$, unless $L=2$. At length two only one diagram contributes, and all subsequent formulas ought to be divided by $2$ at $L=2$.

Any VBS is integrable in the Heisenberg model and so is the boundary state defined by (\ref{uudd}). Its overlaps with the Bethe states can be obtained by taking the isotropic limit of the general XXZ formula \cite{Pozsgay:2018ixm}, but we find it more convenient to first bring the boundary state to the generalized dimer form and then use a simple relation between dimer and N\'eel overlaps \cite{Pozsgay:2009,Pozsgay:2018ixm}. The $SU(2)$ symmetry acts on the elementary block of the boundary state as $K\rightarrow \Omega K\Omega ^t$ and by $\widetilde{K}\rightarrow \Omega \widetilde{K}\Omega ^{-1}$ on the cross-channel reflection matrix $\widetilde{K}=-iK\sigma ^2$. Rescalings $K\rightarrow cK$ only change the  overall normalization. The overlaps thus depend only on the ratio of eigenvalues of $\widetilde{K}$. In our case, $\widetilde{K}=-i\sigma ^2$
 but the overlaps will be the same for $\widetilde{K}'=\sigma ^3$ or
$\left\langle K'\right|=\left\langle \uparrow\downarrow\right|
 +\left\langle \downarrow\uparrow\right|$, which is a generalized dimer.
From  \cite{Pozsgay:2009,Pozsgay:2018ixm} and using the N\'eel-MPS$_2$ equivalence we get:
\begin{equation}
 \left\langle {\rm VBS}_{K}\right.\!\left| \mathbf{u}\right\rangle=
 2^{L-1}\,\frac{\left\langle {\rm MPS}_2\right.\!\left| \mathbf{u}\right\rangle}{Q\left(\frac{i}{2}\right)}\,,
\end{equation}
where the MPS overlap is given by (\ref{MPSk-u}) with $k=2$.

Collecting the pieces we find for the one-point function:
\begin{equation}\label{Cu-su2}
 C_{\mathbf{u}}
 =
 2^{1-L}L^{-\frac{1}{2}}
 \sqrt{\frac{Q\left(0\right)}{Q\left(\frac{i}{2}\right)}\,\,\frac{\det G^+}{\det G^-}}\,,
\end{equation}
in complete agreement with (\ref{structure-const}), (\ref{B1})! For $L=2$ the result should be divided by $2$. To understand better the connection to asymptotic all-loop formulas we pause  to consider BPS operators whose dCFT one-point functions were recently calculated exactly using supersymmetric localization \cite{Komatsu:2020sup}.

\subsection{Protected operators and wrapping}

The BPS operator
\begin{equation}\label{OBPS}
 \mathcal{O}_{\rm BPS}=\left(\frac{4\pi ^2}{\lambda }\right)^{\frac{L}{2}}L^{-\frac{1}{2}}\mathop{\mathrm{tr}}Z^L,
\end{equation}
is an empty vacuum with no Bethe roots. One may expect to find a trivial one-point function due to supersymmetry protection, but the all-loop formula  (\ref{MPSk-u}), (\ref{Tk}) retains some coupling dependence even if all $\det$'s and $Q$'s are set to one.

The supersymmetry protection is thus not complete, it does not eliminate all quantum corrections but restricts them a lot. The BPS one-point functions, as a result, can be computed by localization on hemisphere \cite{Robinson:2017sup,Wang:2020seq,Komatsu:2020sup} and by solving the resulting matrix model at large-$N$ \cite{Wang:2020seq,Komatsu:2020sup}. Slightly changing the notations compared to \cite{Komatsu:2020sup}, we write their result as
\begin{eqnarray}\label{CBPS}
 C_{\rm BPS}&=&2^{-L}L^{-\frac{1}{2}}\left\{
 \left(\frac{16\pi ^2}{\lambda }\right)^{\frac{L}{2}}
 \sum_{a=-\frac{k-1}{2}}^{\frac{k-1}{2}}x_a^L
 -k\delta _{L,2}
 \right.
\nonumber \\
 &&\left. \vphantom{ \sum_{a=-\frac{k-1}{2}}^{\frac{k-1}{2}}x_a^L}
-\frac{\sqrt{\lambda }}{4}\oint\frac{dx}{2\pi i}\,\,\left(1-\frac{1}{x^2}\right)
\frac{1}{\left(ix\right)^L}\,
\kth\left[\frac{\sqrt{\lambda }}{4}\left(x+\frac{1}{x}\right)\right]
\right\},
\end{eqnarray}
where integration is along the unit circle and $\kth$ denotes
\begin{equation}
 \kth =
\begin{cases}
\tanh  & {\rm for~}k {\rm ~even} 
\\
 \coth  & {\rm for~}k{\rm ~odd},
\end{cases}
\qquad 
\ktg=
\begin{cases}
 \tan & {\rm for~}k {\rm ~even} 
\\
 -\cot & {\rm for~}k{\rm ~odd}.
\end{cases}
\end{equation}

This formula is large-$N$ exact and is fully non-perturbative in  the 't~Hooft coupling. The two terms in (\ref{CBPS})  have a distinct origin and  a different interpretation. The one explicitly displayed agrees with the asymptotic integrability formula apart from the $a=0$ term. The remainder was found exponentially suppressed by the operator length in a number of limiting cases \cite{Komatsu:2020sup}, and is naturally interpreted as a wrapping effect in the spin chain or on  the string worldsheet\footnote{We would like to thank Shota Komatsu for clarification of this point.}.

Before setting $k=1$ in the exact formula, we rewrite the integral term in a slightly different form. Changing the integration variable to
\begin{equation}
 \frac{4\pi iu}{\sqrt{\lambda }}=x+\frac{1}{x}\,,
\end{equation}
we find:
\begin{equation}
 \wrap{
=\left(\frac{\lambda }{16\pi ^2}\right)^{\frac{L}{2}}\oint
\frac{du}{2ix_u^L} \,\,\ktg\pi u,
 }
\end{equation}
where $x_u$ is the Zhukovsky variable (\ref{Zhuk}) and the contour now encircles the cut implicit in its definition. Inflating the contour to wrap the poles of $\ktg$ we finally get:
\begin{equation}
  \wrap{
=\left(\frac{\lambda }{16\pi ^2}\right)^{\frac{L}{2}}
\sum_{m\in \mathbbm{Z}+\frac{k-1}{2}}^{}\frac{1}{x_m^L}\,,
 }
\end{equation}
so that the whole answer becomes
\begin{equation}
 C_{\rm BPS}=2^{-L}L^{-\frac{1}{2}}\left[
 \left(\frac{16\pi ^2}{\lambda }\right)^{\frac{L}{2}}
 \sum_{a=-\frac{k-1}{2}}^{\frac{k-1}{2}}x_a^L
 {-k\delta _{L,2}+
 \left(\frac{\lambda }{16\pi ^2}\right)^{\frac{L}{2}}
\sum_{b\in \mathbbm{Z}+\frac{k-1}{2}}^{}\frac{1}{x_b^L}
}
 \right].
\end{equation}

The infinite sum depends on $k$ only through the labels in summation, which are integer or half-integer depending on the parity of $k$. This actually makes a big difference. If $k$ is odd the sum is bosonic (goes over integers) and contains a zero mode. It is natural to consider the zero mode a part of the asymptotic contribution\footnote{This was suggested to us by Shota Komatsu.}, rather than the infinite sum over wrappings. Taking into account that $x_{+0}x_{-0}=-\lambda /16\pi ^2$, and that the spin-chain length is always even, we find:
\begin{equation}
 C_{\rm BPS}=2^{-L}L^{-\frac{1}{2}}\left[
 \left(\frac{16\pi ^2}{\lambda }\right)^{\frac{L}{2}}
 \sum_{a}^{}x_a^L
 {-k\delta _{L,2}+
 \left(\frac{\lambda }{16\pi ^2}\right)^{\frac{L}{2}}
\sum'_{b\in \mathbbm{Z}+\frac{k-1}{2}}\frac{1}{x_b^L}
}
 \right], 
\end{equation}
where the summation over $a$ now goes in the same range as in (\ref{all-loop-Tk}), counting $a=0$ twice, and the summation over $b$ excludes the $b=0$ term. This neatly separates the asymptotic and wrapping effects, and for $k=1$ gives:
\begin{equation}
 C_{\rm BPS}^{k=1}
 \stackrel{\lambda \rightarrow 0}{=}2^{-L}L^{-\frac{1}{2}}\left(
 1
 {+1-\delta _{L,2}}
 \right),
\end{equation}
reproducing the simple combinatorics of the leading-color diagrams from the previous section. 

\subsection{Full $SO(6)$ \label{SO(6)}}

The diagrammatic simplicity of the one-point functions at $k=1$ suggests to look at other types of operators. Extension to all scalars comes at no extra cost. The boundary state in the $SO(6)$ sector is an integrable VBS with
\begin{equation}
 K=ZZ+XX+YY+\bar{Z}\bar{Z}+\bar{X}\bar{X}+\bar{Y}\bar{Y},
\end{equation}
or, in the real scalar basis,
\begin{equation}
 \left\langle K\right|=2\sum_{i=1}^{3}\left\langle ii\right|-2\sum_{j=4}^{6}\left\langle jj\right|.
\end{equation}
An overlap formula for this boundary state was reported in \cite{deLeeuw:2019ebw}, allowing us to immediately write down the determinant representation for the one-point function.

The $SO(6)$ Bethe equations are 
\begin{equation}\label{CartanBAE}
 \,{\rm e}\,^{i\chi _{aj}}\equiv \left(\frac{u_{aj}-\frac{iq_a}{2}}{u_{aj}+\frac{iq_a}{2}}\right)^L\prod_{bk}^{}\frac{u_{aj}-u_{bk}+\frac{iM_{ab}}{2}}{u_{aj}-u_{bk}-\frac{iM_{ab}}{2}}=-1.
\end{equation}
with the Cartan matrix and weight vector
\begin{equation}
 M=\begin{bmatrix}
  2 & -1 & 0\\ 
  -1 & 2 & -1  \\
  0 & -1 & 2 \\
 \end{bmatrix},\qquad 
 q=\begin{bmatrix}
 0  \\ 
  1  \\ 
  0  \\
 \end{bmatrix}.
\end{equation}
Doubling of the fields in the boundary state requires the number of roots of each type to be even, and those have to be paired: $\mathbf{u}=\left\{u_{aj},-u_{aj}\right\}$, to meet the integrability condition. The Gaudin matrix then factorizes\footnote{We review factorization of the Gaudin matrix for Cartan-type Bethe equations in appendix~\ref{FF}.},   and the overlap formula takes literally the same form (\ref{Cu-su2}) as in the $\mathfrak{su}(2)$ case,
where $Q(u)$ should be understood as the complete Baxter polynomial $Q(u)=\prod\limits_{ja}^{}(u^2-u_{aj}^2)$.

\section{Gluons\label{Gluons}}

Having set the stage, it is now time to make good on the promise of going beyond scalar sub-sectors. As a natural intertwiner between the scalar $SU(2)$ sector and the fermionic extension discussed below, we consider the spin-$1$ representation of $SU(2)$. The latter is realized in the gluon sector of the dCFT. To see this, we recall that in ordinary $N=4$ SYM theory the field strength transforms in the reducible representation $(1,0)\oplus(0,1)$ of the Lorentz group. The irreducible components are the self-dual and anti-self-dual parts of $F_{\mu \nu}$
\begin{align}
F^{\mu \nu}=-\tfrac{1}{8} f^{\alpha \beta} (\sigma^{\mu \nu} \epsilon)_{\alpha \beta} - \tfrac{1}{8}  \bar{f}^{\dot\alpha \dot\beta} (\epsilon \bar\sigma^{\mu \nu})_{\dot\alpha \dot\beta} \, ,
\end{align}
which are (anti-)chiral and transform in the spin-$1$ representation of $SU_L(2)$ and $SU_R(2)$,  respectively. For our spinor conventions we refer to Appendix \ref{app:su2spinor}. In the dCFT set-up the original $SU_L(2) \times SU_R(2)$ symmetry is broken to the diagonal subgroup which is equivalent to saying that the four-dimensional Lorentz group $SO(1,3)$ is reduced to the three-dimensional Lorentz group $SO(1,2)$. Obviously, this is due to the fact that Lorentz transformations generated by $L_{\mu 3}$ do not preserve the hyperplane defined by the condition $x_3=0$. In the defect theory it is therefore natural to consider the following linear combinations of the field strength tensor and its Hodge dual
\begin{align}
\label{eqn:SdualFSdCFT}
f^{\hat{\rho}}=F^{3 \hat\rho}+\tfrac{i}{2} \varepsilon^{3 \hat\rho \hat\mu \hat\nu} F_{\hat\mu \hat\nu} \, , \hspace{1.5cm} \bar{f}^{\hat{\rho}}=F^{3 \hat\rho}-\tfrac{i}{2} \varepsilon^{3 \hat\rho \hat\mu \hat\nu} F_{\hat\mu \hat\nu} \, , 
\end{align}
where the hatted index takes values $\hat\rho=0,1,2$. Formally, these combinations can be obtained by multiplying the chiral (anti-chiral) field strength components by  $\bar\sigma^{\hat{\rho}} \sigma^3 \epsilon$ ($\epsilon \sigma^{\hat{\rho}} \bar\sigma^3$) and taking the trace, i.e.
\begin{align}
\label{eqn:SO12SL2Rmap}
f^{\hat{\rho}}=\tfrac{i}{4} f^{\alpha \beta} (\bar\sigma^{\hat{\rho}} \sigma^3 \epsilon)_{\beta \alpha}\, , \hspace{1.5cm} \bar{f}^{\hat{\rho}}=\tfrac{i}{4} \bar{f}^{\dot\alpha \dot\beta} (\epsilon \sigma^{\hat{\rho}} \bar\sigma^3)_{\dot\beta \dot\alpha} \, .
\end{align}
They also correspond to contractions with the 't~Hooft symbol
 \cite{tHooft:1976snw} or, simply speaking, to the space-like $\mathbf{E}\pm \mathbf{B}$ decomposition.

In what follows, we focus on scalar single trace operators composed of $L$ self-dual field strengths
\begin{align}
\label{eqn:SaclarSDualOperator}
\mathcal{O}=\Psi^{\hat\mu_1 \ldots \hat\mu_L} \mathrm{tr} f_{\hat\mu_1} \ldots f_{\hat\mu_L} \, ,
\end{align}
where $\Psi^{\hat\mu_1 \ldots \hat\mu_L}$ is built from three-dimensional metric tensors $\eta^{\hat\mu \hat\nu}$. The perturbative computation of the associated one-point functions pretty much mirrors the situation encountered in the scalar $SU(2)$ sector discussed in section \ref{Scalars}. The leading order term is obtained by just Wick contracting neighboring fields inside the trace. All other contractions are subleading at large-$N$. The propagator of fields in the $N \times N$ block vanishes due to the absence of boundary conditions but the field components in the $\knb$ block acquire a non-vanishing contribution because the gauge field components in this block are subject to Dirichlet/Neumann boundary conditions, cf. table (\ref{bcs}). Imposing Dirichlet boundary conditions on $A_{\hat\rho}$ and Neumann boundary conditions on $A_3$ yields the following propagator for gauge field components in the $\knb$ block
\begin{align}
\langle A_{\mu}^{1a}(x) A_{\nu}^{b1}(y) \rangle = \frac{g_{\rm YM}^2 \delta^{a b}}{8\pi^2} \left(\frac{\eta_{\mu \nu}}{(x-y)^2} - \frac{\eta_{\mu \nu} (-1)^{\delta_{\mu 3}}}{(\bar x-y)^2} \right) \, ,
\end{align}
with no summation implied over $\mu$. The propagator of self-dual field strengths is obtained by neglecting all terms in \eqref{eqn:SdualFSdCFT} which are non-linear in the fields and substituting the above expression for all gauge field contractions. Explicitly, one finds
\begin{align}
\langle f_{\hat\mu}^{1a}(x) f_{\hat\nu}^{b1}(y) \rangle &= - \frac{g_{\rm YM}^2 \delta^{a b}}{\pi^2} \biggl( \frac{\eta_{\hat\mu \hat\nu}}{(\bar x-y)^4} + \frac{8 \eta_{[\hat\mu| [\hat\nu|} (\bar x -y)_{|3]}(\bar x -y)_{|3]}}{(\bar x-y)^6} \nonumber \\
&\hspace{2.3cm} + \frac{2 i \varepsilon_{\hat\mu \hat\nu \hat\kappa 3} (\bar x -y)^{\hat\kappa} (\bar x -y)^{3}}{(\bar x-y)^6} \biggr)  \, ,
\end{align}
where $[]$ denotes antisymmetrization including a factor of $1/2$. For one-point functions only the propagator with coinciding coordinates is relevant which is given by
\begin{align}
\langle f_{\hat\mu}^{1a}(x) f_{\hat\nu}^{b1}(x) \rangle = \frac{g_{\rm YM}^2 \delta^{a b}}{16 \pi^2} \frac{\eta_{\hat\mu \hat\nu}}{x_3^4} \, .
\end{align}

Before we can continue with the computation of one-point functions we need to solve the mixing problem   for operators of the form \eqref{eqn:SaclarSDualOperator}. 
The operators composed of self-dual field strengths comprise the vacuum sector of the $\mathcal{N}=4$ integrable system if the "Beast" grading is used for the Dynkin diagram of $PSU(2,2|4)$ \cite{Beisert:2003yb}. They were studied at length in \cite{Ferretti:2004ba,Beisert:2004fv}.
The mixing matrix in this sector is a Hamiltonian of an integrable spin-$1$ $SU(2)$ spin chain (Zamolodchikov-Fateev model) \cite{Zamolodchikov:1980ku,Kulish:1981gi,Reshetikhin:1986vd}:
\begin{align}
H=\sum\limits_{l=1}^{L} ( 1- P_{l,l+1} + 2 K_{l,l+1}) \, ,
\end{align}
where $P_{l,l+1}$ denotes the permutation operator while $K_{l,l+1}$ denotes the trace operator. 
The model can be solved by Bethe ansatz techniques where each eigenstate $\left|\mathbf{u}\right\rangle$ is characterized by a set of $M$ Bethe roots $\mathbf{u}=\left\{u_1,\ldots ,u_M\right\}$ that satisfy Bethe Ansatz Equations (BAE):
\begin{align}
\label{BAE-su2spin1}
 \,{\rm e}\,^{i\chi_j}\equiv \left(\frac{u_j-i}{u_j+i}\right)^L
 \prod_{k\neq j}^{}\frac{u_j-u_k+i}{u_j-u_k-i}=1 \, ,
\end{align}
with the energy and momentum given by
\begin{align}
\label{EP-su2zspin1}
 E_{\mathbf{u}}=\sum_{j}^{}\frac{4}{u_j^2+1}\,,\qquad 
 \,{\rm e}\,^{iP_{\mathbf{u}}}=\prod_{j}^{}\frac{u_j+i}{u_j- i}\,.
\end{align}
The Bethe ansatz describes the eigenstates of the spin chain in terms of excitations around the ferromagnetic vacuum. The vacuum state has spin $S=L$ and each magnon reduces the spin by $1$. In general, the spin is therefore given by $S=L-M$ and we will focus on the sector $M=L$ with $L$ being even because only scalar operators can have a non-vanishing one-point function. Furthermore, it suffices to concentrate on states for which the rapidities are balanced $\mathbf{u}=\left\{u_j,-u_j\right\}_{j=1\ldots L/2}$ because unbalanced states carry a non-vanishing charge $Q_3$ and therefore have zero overlap with the yet to be defined boundary state. Such states automatically fulfill the zero momentum condition and are thus compatible with the cyclicity of the trace.

Finally, let us now address the definition of the boundary state and present a closed formula for the overlap with Bethe eigenstates. Above we have argued that the leading order term contributing to $\langle \mathcal{O}(x) \rangle$ is obtained by contracting neighboring fields inside the trace. The appropriate boundary state is thus just the $L/2$-fold tensor product of the single two-site state $\Psi_{\rm{sing}}^{\hat\mu \hat\nu}=\eta^{\hat\mu \hat\nu}$:
\begin{align}
\left\langle B\right|=\left\langle \Psi_{\rm{sing}} \right|^{\otimes \frac{L}{2}} \, ,\hspace{2cm}
 B^{\hat\mu_1 \ldots \hat\mu_L}=\eta^{\hat\mu_1 \hat\mu_2} \ldots \eta^{\hat\mu_{L-1} \hat\mu_L} \, .
\end{align}

We are not aware of any exact formulas for overlaps in the Fateev-Zamolodchikov model\footnote{A closely related question of post-quench dynamics in this model was studied in \cite{piroli2016exact}.}, but we can work by analogy. The tensor-product state above projects spins on adjacent sites onto a singlet. A spin-1/2 counterpart would be the dimer state with the two-site block $K=\uparrow\downarrow-\downarrow\uparrow$, whose overlaps with the Bethe states are known. We conjecture that the spin-1 formula is the same up to obvious changes in the Gaudin matrix. In other words, the overlap of the state $\left\langle B\right|$ with a paired spin-$0$ Bethe eigenstate $\left|\mathbf{u} \right\rangle=|u_j,-u_j \rangle$ where $j=1\ldots L/2$ takes the remarkably simple determinant form
\begin{align}
\label{spin-1overlap}
 \frac{\left\langle B \right.\!\!\left|\mathbf{u} \right\rangle}{\left\langle \mathbf{u}\right.\!\!\left| \mathbf{u}\right\rangle^{\frac{1}{2}}}
 = (-2)^{-\frac{L}{2}}\sqrt{\frac{1}{Q\left(0\right)Q(\tfrac{i}{2})}\,\,
 \frac{\det G^+}{\det G^-}
 }\,,
\end{align}
where $Q(u)$ is the Baxter polynomial and $G^\pm$ are $L/2\times L/2$ matrices defined as
\begin{align}
\label{GpmS=1}
G^\pm_{jk}=K^\pm_{jk}+\delta _{jk}\left(\frac{2L}{u_j^2+1}-\sum_{l}^{}K^+_{jl}\right) \, ,
\end{align}
with
\begin{align}
\label{Kpm}
K^\pm_{jk}=\frac{2}{(u_j-u_k)^2+1}\pm\frac{2}{(u_j+u_k)^2+1}\,.
\end{align}
We checked the validity of the above formula for scalar states up to and including $L=10$. Note that in order to recover the structure constants $C_{\mathbf{u}}$ one still needs to insert an additional factor of $2$ as the boundary state effectively picks one out of the two equivalent contractions, cf. section \ref{Scalars}.

The Heisenberg model can be generalized to spins in an arbitrary $SU(2)$ representation. We comment on the spin-$S$ overlap formula appendix~\ref{app:spinS}.

\section{Fermions \label{Fermions}}
It is well-known that the simplest sub-sector involving fermions in ${\cal N}=4$ SYM is the $SU(2|3)$ sub-sector 
containing operators which
are built from the three complex scalars $Z$, $X$ and $Y$ as well as two fermions $\Psi_1$ and $\Psi_2$~\cite{Beisert:2003ys}.  
With the boundary
conditions~(\ref{bcs}) the contraction rules for the components of the fermionic fields relevant in the large-$N$ limit  read
\begin{equation}
\langle \Psi_{\alpha}^{1a}(x) \Psi_{\beta}^{b1}(y) \rangle = \frac{g_{\rm YM}^2}{8\pi^2}\, \epsilon_{\alpha \beta}\,  \delta^{ab} \,
\cdot\frac{\bar{x}_3-y_3}{|\bar{x}-y|^4}.
\end{equation}
This expression is most easily derived by formally taking the $k\rightarrow 1$ limit of 
the AdS propagator relevant for the Dirac fermions for $k>1$ given in~\cite{Buhl-Mortensen:2016jqo} which results in
 \begin{equation}
D_F(x,y)=i\gamma^{\mu} \partial_\mu^x\left(\frac{1}{2}D_+(x,y) (1+i\gamma^3)+\frac{1}{2}D_-(x,y)(1-i\gamma^3).
\right),
\end{equation}
Here we will restrict ourselves to considering the lowest loop level where the dilatation operator is given by eqn~(\ref{Heisenberg}) with $P$ being replaced by the graded permutation. 
For simplicity we will restrict ourselves to the case of one scalar field only. In that case the relevant boundary state for the
computation of one-point functions in the large-$N$ limit is
\begin{align}
 \left\langle B\right|=\bigl(\left\langle ZZ \right| +\left\langle \uparrow \downarrow \right| - \left\langle\downarrow\uparrow  \right| \bigr) ^{\otimes \frac{L}{2}} \, , \label{FBoundary}
\end{align}
where we have represented the two fermions by up and down arrows.  One can show that the boundary state~(\ref{FBoundary}) is 
annihilated by the first odd charge $Q_3$, defined with the graded permutation replacing the usual permutation. This is a simple
first indication that a closed overlap formula should exist. 
It is straightforward to write down an expression for a two fermion eigenstate of the dilatation operator
\begin{equation}
 \mathcal{O}_p=\sum_{l=1}^{L-1} \varepsilon ^{\alpha \beta }\mathop{\mathrm{tr}}\Psi _{\alpha} Z^{l-1}\Psi _{\beta }Z^{L-l-1}\cos p\left(l-\frac{1}{2}\right) \, ,
\end{equation}
with the momenta and the energies  given as
\begin{equation}\label{Q2f}
p_n=\frac{2\pi n}{L-1}, \hspace{0.5cm} n=0,1,\ldots,L-2,
\end{equation}
\begin{equation}\label{E2f}
E_n=8 \sin^2\left(\frac{p_n}{2}\right).
\end{equation}
The momentum quantization condition follows from the symmetry under $l\rightarrow L-l$.  We can readily evaluate the
one-point function corresponding to the operator $\mathcal{O}_p$ and find (up to an irrelevant phase factor which is not determined)
\begin{align}\label{2mag-explicit}
\frac{ \left\langle B  \bigl| \mathcal{O}_p \right\rangle}{\left\langle \mathcal{O}_p  \bigl| \mathcal{O}_p \right\rangle^{1/2}}=\sqrt{\frac{2L}{(L-1)} \frac{u^2}{u^2+\frac{1}{4}}} \, ,
\end{align}
where we have introduced $u=\frac{1}{2}\cot \left(\frac{p}{2}\right)$.
In order to analyze the overlaps for more general operators we write down the Bethe equations for the $SU(2|1)$ spin chain
corresponding to the grading X---O. 
\begin{align}
\label{eqn:BAE1S2F}
1&= \left(\frac{u_k-\frac{i}{2}}{u_k+\frac{i}{2}}\right)^{L} \,
\prod_{l=1}^{{K}^{II}} \frac{u_k-{y}_l+\frac{i}{2}}{u_k-{y}_l-\frac{i}{2}} \, , \nonumber \\
1&=\prod_{l=1}^{K^I} \,\,
\frac{{y}_k-u_l+\frac{i}{2}}{{y}_k-u_l-\frac{i}{2}} 
\prod_{l\neq k}^{{K}^{II}} \,\,
\frac{{y}_k-{y}_l-i}{{y}_k-{y}_l+i} \, .
\end{align}
With this grading we have chosen the vacuum to correspond to the state built entirely
from the bosonic $Z$-fields.  The momentum carrying roots $\{u_i\}$ are fermionic and create fermionic fields of one type, say 
$\Psi_1$, on top of the vacuum. The other roots $\{y_i\}$ are bosonic and change a fermionic excitation of type
$\Psi_1$ into a fermionic excitation of type $\Psi_2$.  In terms of mode numbers the field content of an operator is thus given 
as follows
\begin{equation}
\# Z= L-K_I, \hspace{0.5cm} \# \Psi_1= K_I-K_{II}, \hspace{0.5cm} \# \Psi_2=K_{II}.
\end{equation}
In order for the overlap between a Bethe eigenstate and the boundary state (\ref{FBoundary}) 
to be non-zero, the following selection rules must be fulfilled
\begin{equation}
L, K_I \hspace{0.5cm} \mbox{even}, \hspace{0.5cm} K_{II}=K_I/2. 
\end{equation} 
Given that the charge $Q_3$ annihilates the boundary state (\ref{FBoundary})  (and assuming the same to be the case for 
all higher odd charges) the roots of type $u_i$ have to come in pairs of opposite signs.
If $K_{II}$ is even the same is the case for the roots ${y_i}$, and if $K_{II}$ is odd there will be an additional single root at zero.
The precise argument for this is a copy of the corresponding argument given for the $SU(3)$ spin chain in~\cite{deLeeuw:2016umh}.

As a warm up let us recover the two-excitation state from above from the Bethe equations.  This state has quantum numbers
$K_I=2$ and $K_{II}=1$, and the corresponding Bethe roots have to take the form
\begin{equation}
u_1=-u_2=u, \hspace{0.5cm} y_1=0.
\end{equation}
This trivializes the momentum constraint as well as the second Bethe equation whereas the first one is reduced to
\begin{align}
1=\left(\frac{u-\frac{i}{2}}{u+\frac{i}{2}}\right)^{L-1} \equiv e^{i p (L-1)} \, ,
\end{align}
which exactly reproduces the results for momenta and energies given above, cf.\ eqns.~(\ref{Q2f}) and~(\ref{E2f}).

Moving on to higher excited states as usual requires numerical means. 
We have generated eigenstates by explicit diagonalization and computed Bethe roots numerically. In this way we have been able to compute the overlaps for states up to length 10 and we find that they are all expressible in the general form
\begin{equation}
\label{general-formula}
\frac{ \left\langle B  | {\bf u}, {\bf y}  \right\rangle}{\left\langle \bf{u}, \bf{y}  |  \bf{u}, \bf{y} \right\rangle^{1/2}}=
\sqrt{\frac{Q_u(0)} {\bar{Q}_y(0)\bar{Q}_y\left(\frac{i}{2}\right) }\,\frac{\det G_+}{\det G_-}}\, ,
\end{equation}
where the bar means that roots at zero have to be excluded from the product.  We elaborate on the precise factorization
form of the Gaudin matrix in appendix~\ref{FF}.

\section{Conclusions\label{Conclusion}}

The one-point function in the D3-D5 dCFT simplify at $k=1$ retaining all their integrability properties. The associated boundary state is of conventional valence bond type, in contradistinction to less common matrix product states arising at $k>1$. These simplifications allowed us to find overlap formulas for operators with gluon and fermion constituents, which has proven prohibitively complicated at $k>1$. 

The $k=1$ case is also special from the bootstrap perspective. The magnons of the spin chain (or string modes in AdS) can form bound states with the defect in the cross channel, there are exactly $k$ such states on the defect with $k$ units of flux \cite{Komatsu:2020sup}. For $k=1$ the bound states are obviously absent.

Our explicit diagrammatic calculations perfectly agree with the asymptotic all-loop formula for one-point functions, thus testing it at order $\mathcal{O}(\lambda ^{L/2})$ of perturbation theory. We believe that the $k=1$ dCFT, due to its intrinsic simplicity, is the best playground for TBA-type generalizations. 

\subsection*{Acknowledgements}
We would like to thank S.~Frolov, A.~Ipsen, S.~Komatsu and D.~Volin for interesting discussions. The work of KZ was supported by the grant "Exact Results in Gauge and String Theories" from the Knut and Alice Wallenberg foundation and by RFBR grant 18-01-00460 A. 
The work of C.\ Kristjansen and D.\ M\"{u}ller was supported by DFF-FNU through grant number DFF-4002-00037.

\appendix

\section{Spinor Conventions \label{app:su2spinor}}
In this appendix we present our conventions for $SU(2)$ spinors. Unless stated otherwise, we work in Minkowski space with the metric given by $\eta_{\mu \nu}=\mathrm{diag}(+1,-1,-1,-1)$. We begin by introducing the four-dimensional sigma matrices
\begin{align}
\sigma^{\mu  \da \b}=(\mathds{1},\vec{\sigma}) \, ,  \hspace{2cm}  \bar{\sigma}^{\mu}_{ \a\db}=(\mathds{1},-\vec{\sigma}) \, ,
\label{eqn: 4dsimga}
\end{align} 
where $\mathds{1}$ denotes the identity matrix and $\vec{\sigma}$ denotes the $3$-vector of Pauli matrices
\begin{align}
\sigma^1 = \begin{pmatrix} 0 & 1 \\ 1 & 0 \end{pmatrix}\, , \qquad \sigma^2 = \begin{pmatrix} 0 & -i \\ i & 0 \end{pmatrix} \, , \qquad \sigma^3 = \begin{pmatrix} 1 & 0 \\ 0 & -1  \end{pmatrix} \, .
\end{align} 
$SU(2)$ indices are raised and lowered with the help of the completely antisymmetric two-tensors
\begin{align}
\epsilon^{\a \b}=\epsilon_{\a \b}=\begin{pmatrix}
0  & 1 \\
-1 & 0
\end{pmatrix} \, , \hspace{2cm} 
\epsilon^{\da \db}=\epsilon_{\da \db}=\begin{pmatrix}
0  & -1 \\
1 & 0
\end{pmatrix} \, .
\end{align}
Furthermore, we define the following antisymmetric combinations of four-dimensional sigma matrices
\begin{align}
\sigma\indices{^{\mu \nu} _{ \a} ^{\b}}:=\tfrac{i}{2} \bigl(\bar\sigma^{\mu}_{ \a \dg} \, \sigma \indices{^{\nu \dg \b}}-\bar\sigma^{\nu}_{ \a \dg} \, \sigma \indices{^{\mu \dg \b}} \bigr)\, , \qquad \bar\sigma \indices{^{\mu \nu \da} _{\db}}:=\tfrac{i}{2} \bigl(\sigma \indices{^{\mu \da \g}} \, \bar\sigma^{\nu}_{ \g \db}-\sigma \indices{^{\nu \da \g}} \, \bar\sigma^{\mu}_{  \g \db} \bigr) \, ,
\label{eqn: antisymsigma}
\end{align}   
which allow us to assign two bispinors to an antisymmetric $2$-tensor $F^{\mu \nu}$ as follows
\begin{align}
f^{\a \b}:=F_{\mu \nu} \, (\epsilon\sigma^{\mu \nu})^{\a \b} \, , \hspace{2cm}
\bar{f}^{\da \db}:=F_{\mu \nu} \, (\bar{\sigma}^{\mu \nu} \epsilon)^{\da \db} \, .
\end{align}
The following trace identities come in handy when checking \eqref{eqn:SO12SL2Rmap}
\begin{align}
\Tr \left(\bar{\sigma}^{\mu} \, \sigma^{\nu}\right) &=2 \, \eta^{\mu \nu} \, , \nn \\
\Tr \left(\bar\sigma^{\mu}\, \sigma^{\nu} \, \bar\sigma^{\rho} \, \sigma^{\kappa} \right) &= 2 \left(
\eta^{\mu\nu}\, \eta^{\rho\kappa} + \eta^{\nu\rho}\, \eta^{\mu\kappa} -
\eta^{\mu\rho}\, \eta^{\nu\kappa} -i\,\varepsilon^{\mu\nu\rho\kappa} \right)  \, , \nn \\
\Tr \left( \sigma^{\mu} \, \bar\sigma^{\nu}\, \sigma^{\rho}\,\bar\sigma^{\kappa} \right) &=2 \left(
\eta^{\mu\nu}\, \eta^{\rho\kappa} + \eta^{\nu\rho}\, \eta^{\mu\kappa} -
\eta^{\mu\rho}\, \eta^{\nu\kappa} +i\,\varepsilon^{\mu\nu\rho\kappa} \right) \, .
\label{eqn: sigmatrace}
\end{align}
Finally, we state our conventions for the four-dimensional gamma matrices
\begin{align}
\gamma^\mu=\begin{pmatrix}
0  &  \sigma^{\mu}\\
\bar{\sigma}^\mu & 0
\end{pmatrix} \, , \hspace{2cm} \left\{\gamma^\mu, \gamma^\nu \right\}=2 \, \eta^{\mu \nu} \,.
\label{eqn: cl13}
\end{align}

\section{Factorization Formulae \label{FF}}

The Gaudin matrix for general Cartan-type Bethe equations (\ref{CartanBAE}) is
\begin{equation}
 G_{aj,bk}\equiv \frac{\partial \Phi _{aj}}{\partial u_{bk}}
 =\left(\frac{Lq_a}{u_{aj}^2+\frac{1}{4}}-\sum_{cl}^{}K_{aj,cl}\right)\delta _{ab}\delta _{jk}+K_{aj,bk},
\end{equation}
where
\begin{equation}
  K_{aj,bk}=\frac{M_{ab}}{(u_{aj}-u_{bk})^2+\frac{M_{ab}^2}{4}}\,.
\end{equation}
Suppose that roots are fully paired. Then using the determinant identity
\begin{equation}
 \det\begin{bmatrix}
  A & B \\ 
  B & A \\ 
 \end{bmatrix}=\det(A-B)\det(A+B),
\end{equation}
the Gaudin determinant factorizes as
\begin{equation}
 \det G=\det G^+\det G^-
\end{equation}
with
\begin{equation}
 G^\pm_{aj,bk}=\left(\frac{Lq_a}{u_{aj}^2+\frac{1}{4}}-\sum_{cl}^{}K^+_{aj,cl}\right)\delta _{ab}\delta _{jk}+K^\pm_{aj,bk},
\end{equation}
where
\begin{equation}
  K_{aj,bk}^\pm=\frac{M_{ab}}{(u_{aj}-u_{bk})^2+\frac{M_{ab}^2}{4}}\pm\frac{M_{ab}}{(u_{aj}+u_{bk})^2+\frac{M_{ab}^2}{4}}\,.
\end{equation}

In the presence of zero roots factorization formulae are modified. Suppose there are zero roots at levels $a_1,\ldots ,a_n$, namely
\begin{equation}
 u_{a_\alpha 0}=0,\qquad \alpha =1\ldots n.
\end{equation}
The determinant formula to use is
\begin{equation}
 \det\begin{bmatrix}
  A & B  & v \\ 
  B & A & v \\ 
  v^t & v^t  & g \\ 
 \end{bmatrix}=\det(A-B)\det\begin{bmatrix}
 A+B  & \sqrt{2}\,v \\ 
  \sqrt{2}\,v^t & g \\ 
 \end{bmatrix},
\end{equation}
valid for any square matrices $A$ and $B$ of size $N\times N$, $N\times n$ matrix $v$ and $n\times n$ matrix $g$. 

As a result, both Gaudin factors are modified to include zero roots:
\begin{equation}
 G^\pm_{aj,bk}=\left(\frac{Lq_a}{u_{aj}^2+\frac{1}{4}}-\sum_{cl}^{}K^+_{aj,cl}
 -\frac{1}{2}\sum_{\alpha }^{}\,K^+_{aj,a_\alpha 0}\right)\delta _{ab}\delta _{jk}+K^\pm_{aj,bk},
\end{equation}
where indices $aj$ etc run over positive paired roots. The $G^+$ matrix acquires $n$ additional rows and columns:
\begin{eqnarray}
 G^+_{aj,\alpha }&=&\frac{1}{\sqrt{2}}\,K^+_{aj,a_\alpha 0}
\nonumber \\
 G^+_{\alpha \beta }&=&\left(\frac{4L}{q_{a_\alpha }}-\sum_{cl}K^+_{a_\alpha 0,cl}-\sum_{\gamma }^{}\frac{4}{M_{a_\alpha a_\gamma }}\right)
 \delta _{\alpha \beta }+\frac{4}{M_{a_\alpha b_\beta }}\,.
\end{eqnarray}
In the last formula, $1/q_{a_\alpha }\rightarrow 0$ if $q_{a_\alpha }=0$ and the same for $1/M_{a_\alpha a_\beta }$.

\section{$SU(2)$ overlaps for arbitrary spin\label{app:spinS}}

The integrable $\mathfrak{su}(2)$ spin chain with spins in the $(2S+1)$-dimensional representation (Takhtajan-Babujian model \cite{A.:1982zz,Babujian:1982ib,Babujian:1983ae}) is defined by the Hamiltonian:
\begin{equation}
 H=\sum_{l=1}^{L}\sum_{j=0}^{2S}\psi (2j+1)P_{l,l+1}^j \, ,
\end{equation}
where $P^j_{l,l+1}$ is the projector on the spin-$j$ component in the tensor product decomposition $[S]_l\otimes [S]_{l+1}=\bigoplus\limits_{j=0}^{2S}[j]$ and $\psi (n)$ is the harmonic number. 
The Bethe Ansatz Equations for the model are
\begin{equation}
 \left(\frac{u_j+iS}{u_j-iS}\right)^L=-\prod_{k}^{}\frac{u_j-u_k+i}{u_j-u_k-i}\,.
\end{equation}
The natural generalization of the singlet state that we consider in sec.~\ref{Gluons} is the VBS with
\begin{equation}
 K_{l,l+1}=P_{l,l+1}^0 \, .
\end{equation}
This state has non-zero overlaps only with paired spin-$0$ eigenstates  $\left|\mathbf{u} \right\rangle=|u_j,-u_j \rangle$, $j=1\ldots S L/2$ and we conjecture these overlaps to be expressible as
\begin{align}
 \frac{\left\langle {\rm VBS}^S\right.\!\!\left|\mathbf{u} \right\rangle}{\left\langle \mathbf{u}\right.\!\!\left| \mathbf{u}\right\rangle^{\frac{1}{2}}}
 =2^{-\frac{L}{2}}\sqrt{ \frac{1}{Q(0)Q(\frac{i}{2})}\,\,
 \frac{\det G^+}{\det G^-}}\,,
\end{align}
where the Gaudin factors are now $SL/2\times SL/2$ matrices:  
\begin{equation}
  G^\pm_{jk}=K^\pm_{jk}+\delta _{jk}\left(\frac{2SL}{u_j^2+S^2}-\sum_{l}^{}K^+_{jl}\right) \, ,
\end{equation}
and $K^\pm$ are given by the same expression (\ref{Kpm}).
For $S=1/2$ this gives the known overlap formula for the dimer \cite{Pozsgay:2018ixm} and for $S=1$ the formula reduces to (\ref{spin-1overlap}).

We also believe that the formula holds for non-compact $\mathfrak{sl}(2)$ spin chains, which corresponding to negative $S$, and for which the singlet-projector VBS is also naturally defined. At $S=-1$, this formula is analogous to the recently derived overlap with the generalized N\'eel state \cite{Jiang:2019xdz,Jiang:2020sdw}, which has the same structure, but involves a different ratio of Baxter polynomials. 

\bibliographystyle{nb}
%\bibliography{refs}

\end{document}